\documentclass{iau}
\usepackage{graphicx}
\usepackage{amsmath}
\usepackage{sidecap}
\usepackage{natbib}

\newcommand{\physrep}{Phys.~Rep.}                   
\newcommand{\apj}{ApJ}                              
\newcommand{\aap}{A~\&~A}                           
\newcommand{\mnras}{MNRAS}                          
\newcommand{\pre}{Phys.~Rev.~E}
\newcommand{\prd}{Phys.~Rev.~D}

\newcommand{\njph}{NJPh}  			    

\title[IAUS 294~~Dynamo \& First Stars] 
{Small-Scale Dynamo Action \\ in Primordial Halos}

\author[Schober et al.]   
{Jennifer~Schober$^1$
 \and Dominik~R.~G.~Schleicher$^2$
 \and Ralf~S.~Klessen$^1$
 \and Christoph~Federrath$^3$
 \and Stefano~Bovino$^2$
 \and Simon~Glover$^1$
 \and Robi~Banerjee$^4$}

\affiliation{
$^1$Heidelberg University, Center for Astronomy, Institute for Theoretical Astrophysics, Albert-Ueberle-Str\,2, 69120 Heidelberg, Germany 

$^2$G{\"o}ttingen University, Institute for Astrophysics, Friedrich-Hund-Platz 1, 37077 G{\"o}ttingen, Germany 

$^3$Monash University, School of Mathematical Sciences, Monash Centre for Astrophysics, Clayton Vic 3800, Australia 

$^4$Hamburger Sternwarte, Gojenbergsweg 112, 21029 Hamburg, Germany
}

\pubyear{2013}
\volume{294}  
\pagerange{119--126}
\jname{Solar and Astrophysical Dynamos and Magnetic Activity}
\editors{A.G. Kosovichev, E.M. de Gouveia Dal Pino \& Y.Yan, eds.}

\begin{document}

\maketitle

\begin{abstract}
The first galaxies form due to gravitational collapse of primordial halos. During this collapse, weak magnetic seed fields get amplified exponentially by the small-scale dynamo - a process converting kinetic energy from turbulence into magnetic energy. We use the Kazantsev theory, which describes the small-scale dynamo analytically, to study magnetic field amplification for different turbulent velocity correlation functions. For incompressible turbulence (Kolmogorov turbulence), we find that the growth rate is proportional to the square root of the hydrodynamic Reynolds number, $\mathrm{Re}^{1/2}$. In the case of highly compressible turbulence (Burgers turbulence) the growth rate increases proportional to $\mathrm{Re}^{1/3}$. With a detailed chemical network we are able to follow the chemical evolution and determine the kinetic and magnetic viscosities (due to Ohmic and ambipolar diffusion) during the collapse of the halo. This way, we can calculate the growth rate of the small-scale dynamo quantitatively and predict the evolution of the small-scale magnetic field. As the magnetic energy is transported to larger scales on the local eddy-timescale, we obtain an estimate for the magnetic field on the Jeans scale. Even there, we find that equipartition with the kinetic energy is reached on small timescales. Dynamically relevant field structures can thus be expected already during the formation of the first objects in the Universe.
\keywords{ISM: magnetic fields, MHD, stars: formation, early universe}
\end{abstract}

\firstsection 

\section{Introduction}

The present-day Universe is strongly magnetized, which is shown by observations of various astrophysical phenomena like jets and outflows caused by magnetic fields. The problem however is that theory predicts the first seed field to be extremely weak. Thus, we need to explain how these fields are amplified and reach the observed present-day large field strengths. \\
A process that can amplify seed fields on very short timescales is the small-scale or turbulent dynamo. The small-scale dynamo converts kinetic energy from turbulence into magnetic energy very efficiently. It strongly depends on the nature of turbulence and here we explore the range reaching from incompressible Kolmogorov turbulence \citep{Kolmogorov1941} to highly compressible Burgers turbulence \citep{Burgers1948}. \\
The first point in time where the dynamo could operate is during the formation of the first stars at a redshift of about 20. Here the turbulence in the primordial halo is generated by accretion \citep{GreifEtAl2008,WiseEtAl2008,LatifEtAl2012} and the weak seed fields created in the very early Universe or by battery processes are amplified by the small-scale dynamo \citep{SchleicherEtAl2010,FederrathEtAl2011.2,SurEtAl2012,PetersEtAl2012}. \\
The outline of this article is as follows: In Sec.~\ref{Basics} we discuss the basic assumptions of our model including the strength of the magnetic seed fields, the description of turbulence and the chemical and thermal evolution of primordial gas during collapse. Furthermore, we calculate the typical MHD quantities like the hydrodynamic and magnetic Reynolds number. Sec.~\ref{KinematicSSD} is on the kinematic growth phase of the small-scale dynamo. We discuss the influence of different types of turbulence on this exponential growth and give the basic steps for deriving the Kazantsev equation, which describes the kinematic dynamo analytically. We derive the critical magnetic Reynolds numbers for small-scale dynamo action and the growth rates for different types of turbulence from the Kazantsev equation. In Sec.~\ref{NonLinearSSD} we describe how the magnetic energy can be transported from the smallest scale of the inertial range to the largest, i.~e.~the nonlinear growth phase. In Sec.~\ref{Conclusion} we summarize our results, which are presented in much more detail in \cite{SchoberEtAl2012.1,SchoberEtAl2012.2,SchoberEtAl2012.3}.

\section{Basics of Our Model}
\label{Basics}

{\underline{\it Magnetic Seed Fields}}. 
Different theories describe the origin of primordial magnetic fields. The first seed fields could already have been produced during inflation. \cite{TurnerWidrow1988} find that $B_0\approx10^{-31}$--$10^{-10}$ G on a scale of $1$ Mpc can be produced when the conformal invariance is broken. Following \cite{Sigl1997}, there is also the possibility of creating a magnetic field during the phase transitions in the very early Universe. They predict a field strength $B_0 \approx 10^{-29}~\mathrm{G}$ from the electroweak phase transition and $B_0 \approx 10^{-20}~\mathrm{G}$ from the quantum chromodynamics (QCD) phase transition. Besides that magnetic fields can be generated by batteries mechanisms like the Biermann battery leading to typical field strengths of $10^{-18}~\mathrm{G}$ \citep{Xu2008}.

{\underline{\it Our Approach to Turbulence}}.
A theoretical description of turbulence starts with the decomposition of the velocity field $\bf{v}$ into a mean field $\left\langle\bf{v}\right\rangle$ and a turbulent component $\delta\bf{v}$:
\begin{equation}
  \bf{v} = \left\langle\bf{v}\right\rangle + \delta\bf{v}.
\end{equation}
The correlation of two turbulent velocity components at the positions $\bf{r}_1$ and $\bf{r}_2$ at the times $t$ and $s$ for a Gaussian random velocity field with zero mean, which is isotropic, homogeneous and $\delta$-correlated in time, is
\begin{equation}
  \left\langle \delta v_i({\bf{r}}_1, t) \delta v_j({\bf{r}}_2, s)\right\rangle =\left[\left(\delta_{ij}-\frac{r_i r_j}{r^2}\right)T_{\mathrm{N}}(r) + \frac{r_ir_j}{r^2}T_{\mathrm{L}}(r)\right]\delta(t-s)
\end{equation}
with $r\equiv|\bf{r}_1-\bf{r}_2|$ and the transverse and longitudinal parts of the two-point correlation function, $T_{\mathrm{N}}$ and $T_{\mathrm{L}}$. Any turbulent flow can in general be described by the relation between the velocity $v(\ell)$ and the size $\ell$ of a velocity fluctuation,
\begin{equation}
  v(\ell) \propto \ell^{\vartheta}.
\label{TurbPower}
\end{equation}
The power-law index $\vartheta$ varies from its minimal value of $\vartheta=1/3$ for Kolmogorov theory \citep{Kolmogorov1941}, i.~e.~incompressible turbulence, to Burgers turbulence \citep{Burgers1948}, i.~e.~highly compressible turbulence, where $\vartheta$ gets its maximal value of $1/2$ \citep{Schmidt2009}. \\
We use the model for the correlation function of the turbulent velocity field that was motivated in \cite{SchoberEtAl2012.1}. The longitudinal correlation function on the different length scales is
\begin{equation}
  T_\mathrm{L}(r) = \begin{cases} 
                      \frac{VL}{3}\left(1-\mathrm{Re}^{(1-\vartheta)/(1+\vartheta)}\left(\frac{r}{L}\right)^{2}\right)     & 0<r<\ell_\nu \\ 
  		      \frac{VL}{3}\left(1-\left(\frac{r}{L}\right)^{\vartheta+1}\right)                                    & \ell_\nu<r<L \\ 
  		      0                                                                                                    & L<r, 
  		    \end{cases}
\label{TL}
\end{equation}
where $\ell_\nu=L~\mathrm{Re}^{-1/(\vartheta+1)}$ denotes the viscous scale of the turbulence and $L$ and $V$ the length and velocity of the largest eddies. The transverse correlation function for the general slope of the turbulent velocity spectrum is
\begin{equation}
  T_\mathrm{N}(r) = \begin{cases} 
                      \frac{VL}{3} \left(1 - \frac{21-38\vartheta}{5} \mathrm{Re}^{(1-\vartheta)/(1+\vartheta)} \left(\frac{r}{L}\right)^{2}\right) & 0<r<\ell_\nu \\
  		      \frac{VL}{3} \left(1 - \frac{21-38\vartheta}{5} \left(\frac{r}{L}\right)^{\vartheta+1}\right)                                 & \ell_\nu<r<L \\ 
  		      0                                                                                                                             & L<r.
  	            \end{cases}
\label{TN}
\end{equation}

{\underline{\it Chemical and Thermal Evolution}}.
We determine the chemical and thermal evolution of gravitationally collapsing primordial gas using the one-zone model of \cite{GloverSavin2009}. We include a modification relating the collapse time to the equation of state \citep{SchleicherEtAl2009} and additional Li and HeH$^+$ chemistry \citep{BovinoEtAl2011.1,BovinoEtAl2011.2}. Our chemical network includes around 30 different atomic and molecular species linked by around 400 different chemical reactions. The initial elemental abundances are primordial \citep{Cyburt2004,GloverSavin2009}, the initial density is  $n_{0} = 1~{\mathrm{cm}^{-3}}$ and the temperature of the gas $T_{0} = 1000$~K, but we have verified that our results are not sensitive to these values.\\
In the one-zone model the mass density $\rho$ evolves as
\begin{equation}
    \frac{\mathrm{d} \rho}{\mathrm{d} t} \propto \frac{\rho}{t_\mathrm{ff}},
\end{equation}
where $t_\mathrm{ff} = \sqrt{3\pi/(32 G \rho)}$ is the free-fall time. Moreover, the temperature evolution is determined by the energy equation,
\begin{equation}
    \frac{\mathrm{d} \epsilon}{\mathrm{d} t} = \frac{p}{\rho^2} \frac{\mathrm{d}\rho}{\mathrm{d} t} - \Lambda_\mathrm{cool} + \Lambda_\mathrm{heat},
\end{equation}
where $\epsilon$ is the specific internal energy, $p$ is the thermal pressure and $\Lambda_\mathrm{cool}$ and $\Lambda_\mathrm{heat}$ are the total cooling and the heating rate per unit mass, respectively. \\
We find that the abundance of H is constant at low densities, but decreases at densities higher than about $10^{10}~\mathrm{cm^{-3}}$ due to the formation of H$_{2}$. For the magnetic properties of the primordial gas the abundances of the charged species are especially important. They determine for example the conductivity of the gas. At densities $n < 10^{8} \: {\mathrm{cm}^{-3}}$, ionized hydrogen is the main positive ion, while at higher densities, Li$^{+}$ dominates.

\begin{figure}
 \includegraphics[width=3.4in]{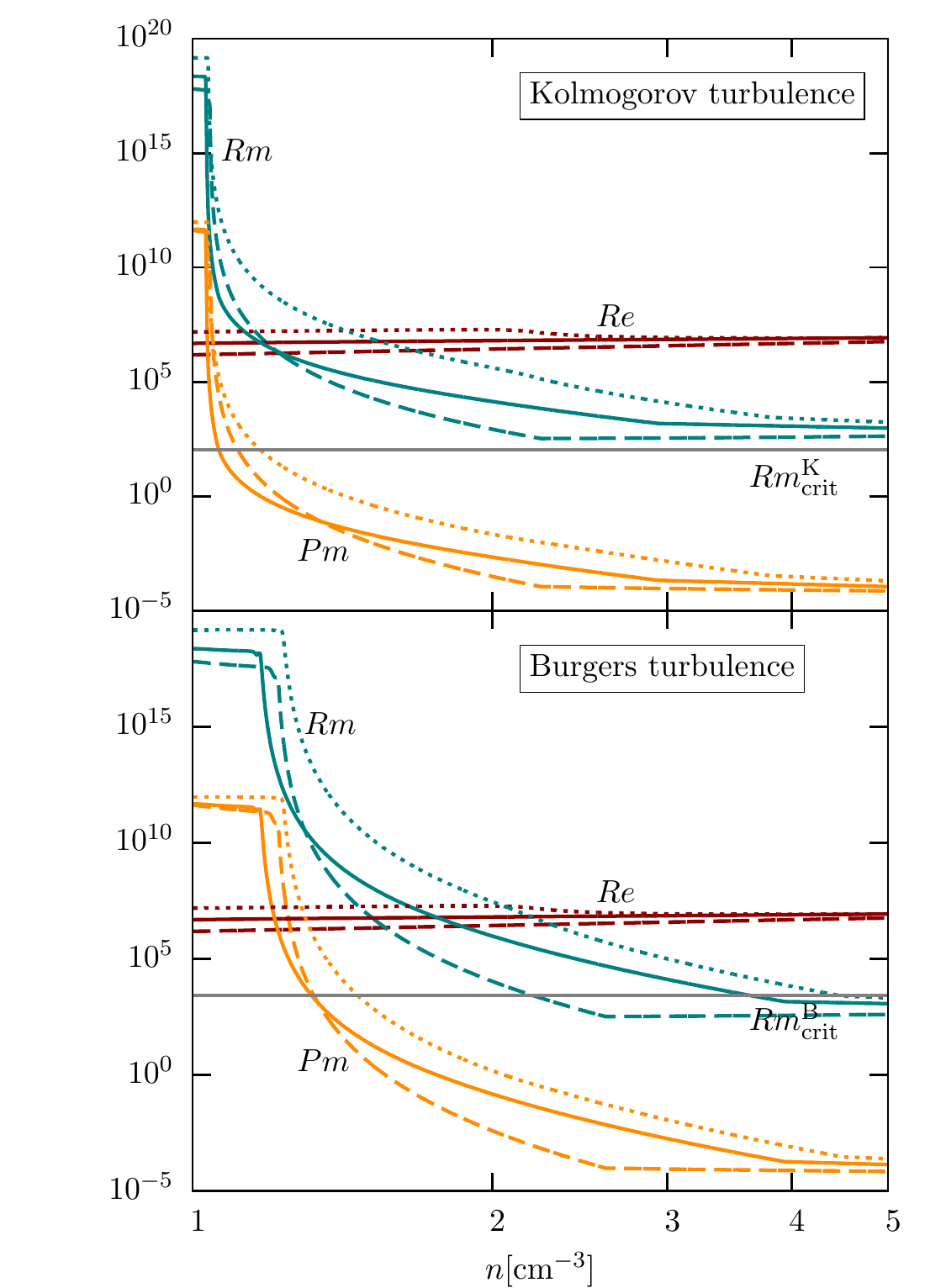} 
 \caption{The hydrodynamic and magnetic Reynolds numbers, Re and Rm, and the magnetic Prandtl number Pm as a function of particle density $n$. The dotted lines correspond to an initial temperature of $10^4$ K, the solid ones to $10^3$ K and the dashed ones to $10^2$ K.}
   \label{Numbers}
\end{figure}

{\underline{\it Characteristic Magnetohydrodynamical Quantities}}.
The kinematic viscosity is 
\begin{equation}
  \nu = \frac{1}{4 d^2 n}\left(\frac{k T}{\pi m}\right)^{1/2},
\label{nu}
\end{equation}
if the molecules are assumed to be rigid spheres \citep{Choudhuri1998}. Here, $n=\rho/m$ is the number density, $d$ the mean particle diameter, $k$ the Boltzmann constant, $T$ the temperature and $m$ the mean molecular weight. \\ 
The two dominant effects that lead to dissipation of magnetic energy are Ohmic resistivity $\eta_{\mathrm{Ohm}}$ and ambipolar diffusion $\eta_{\mathrm{AD}}$, which we calculate according to \cite{PintoEtAl2008}:
\begin{subequations}
\begin{eqnarray}
  \eta_{\mathrm{Ohm},\mathrm{n}} & = & \frac{c^2}{4 \pi \sigma_{||,\mathrm{n}}},  \\
  \eta_{\mathrm{AD},\mathrm{n}} & = & \frac{c^2}{4 \pi} \left(\frac{\sigma_{\mathrm{P,n}}}{\sigma_{\mathrm{P,n}}^2+ \sigma_{\mathrm{H,n}}^2}-\frac{1}{ \sigma_{||,\mathrm{n}}}\right).
\end{eqnarray}
\label{eta}
\end{subequations}
Here $\sigma_{||,\mathrm{n}}$, $\sigma_{P,\mathrm{n}}$ and $\sigma_{H,\mathrm{n}}$ are different conductivities and the index n refers to a neutral species. We focus on the most important neutral species H, He and $\mathrm{H}_2$ and the charged species $\mathrm{H}^+$, $\mathrm{e}^-$ and $\mathrm{Li}^+$. The total Ohmic magnetic diffusivity is $\eta_{\mathrm{Ohm}} = \sum_{\mathrm{n}} \eta_{\mathrm{Ohm,n}}$ and the total resistivity due to ambipolar diffusion is $\eta_{\mathrm{AD}} = 1/(\sum_{\mathrm{n}} \eta_{\mathrm{AD,n}}^{-1})$.\\
The hydrodynamic and magnetic Reynolds numbers are defined as 
\begin{subequations}
\begin{eqnarray}
  \mathrm{Re} & \equiv & \frac{VL}{\nu}, \\
  \mathrm{Rm} & \equiv & \frac{VL}{\eta},
\label{Reynolds}
\end{eqnarray}
\end{subequations}
where $L$ is the length of the largest turbulent fluctuations, i.~e.~the Jeans length, and $V$ the typical velocity on that scale. For the calculation of the magnetic Reynolds number we use the sum of $\eta_\mathrm{Ohm}$ and $\eta_\mathrm{AD}$. \\
The ratio between viscosity and magnetic diffusivity is called magnetic Prandtl number
\begin{equation}
  \mathrm{Pm} \equiv \frac{\mathrm{Rm}}{\mathrm{Re}} = \frac{\nu}{\eta}. 
\label{Pm}
\end{equation}
In Fig.~\ref{Numbers} the density dependency of the Reynolds numbers and the magnetic Prandtl number is shown for both Kolmogorov and Burgers turbulence. We point out that the rapid decrease of Rm and Pm is caused by the dynamo amplification of the magnetic field. In the beginning of the collapse Ohmic resistivity is the dominant diffusion process. With increasing magnetic field $\eta_\mathrm{AD}$ increases proportional to $B^2$ and becomes the main process for magnetic diffusion. Since Rm and Pm are both proportional to $1 / \eta_{\mathrm{AD}}$, in the limit where $\eta_{\mathrm{AD}}\gg\eta_{\mathrm{Ohm}}$, both decrease rapidly with increasing magnetic field strength.\\

\section{Kinematic Small-Scale Dynamo}
\label{KinematicSSD}

{\underline{\it Phenomenology}}.
The small-scale dynamo converts turbulent motions into magnetic energy. An illustrative model describing this process is the stretch-twist-fold dynamo \citep{VainshteinZeldovich1972}. The stretching of a closed magnetic flux rope leads to amplification of the magnetic field strength, as the magnetic flux is a conserved quantity. Afterwards the rope is stretched, twisted and folded such that the original shape is regained. This is an efficient process for increasing the field strength. \\
In the limit of high magnetic Prandtl numbers this mechanism works fastest on the viscous scale $\ell_\nu = \mathrm{Re}^{-1/(\vartheta+1)}~L$, as here the eddy timescale has its minimum. During the transition from large to small magnetic Prandtl numbers the resistive scale $\ell_\eta = \mathrm{Rm}^{-1/(\vartheta+1)}~L$ becomes larger than $\ell_\nu$. Then amplification takes place at roughly $\ell_\eta$, which lies within the inertial range of the turbulent velocity spectrum. Due to larger timescales of the turbulent eddies in the inertial range, we expect the small-scale dynamo to be less efficient at low magnetic Prandtl numbers.

{\underline{\it Kazantsev Theory}}.
Like the velocity field, the magnetic field can be separated into a mean field $\left\langle\textbf{B}\right\rangle$ and a fluctuation part $\delta \textbf{B}$:
\begin{equation}
  \textbf{B} = \left\langle \textbf{B}\right\rangle + \delta \textbf{B}.
\end{equation}
Assuming that the fluctuating component $\delta \textbf{B}$ is a homogeneous, isotropic Gaussian random field with zero mean like the velocity field, we can write down the correlation function as
\begin{equation}
  \left\langle \delta B_i(\textbf{r}_1,t) \delta B_j(\textbf{r}_2,t)\right\rangle = \left(\delta_{ij}-\frac{r_ir_j}{r^2}\right)M_{\mathrm{N}}(r,t) + \frac{r_ir_j}{r^2}M_{\mathrm{L}}(r,t).
\end{equation}
The time derivative of the correlation function is
\begin{eqnarray}
  \frac{\partial}{\partial t} \left\langle \delta B_i \delta B_j\right\rangle = \left\langle\frac{\partial B_i}{\partial t}B_j\right\rangle + \left\langle B_i \frac{\partial B_j}{\partial t}\right\rangle -\frac{\partial}{\partial t}\left(\left\langle B_i\right\rangle\left\langle B_j\right\rangle\right).
\label{DeriMij}
\end{eqnarray}
In the upper equation we can substitute the induction equation
\begin{equation}
  \frac{\partial \textbf{B}}{\partial t} = \mathbf{\nabla}\times\textbf{v}\times\textbf{B} - \eta\mathbf{\nabla}\times\mathbf{\nabla}\times\textbf{B},
\label{induction}
\end{equation}
and the evolution equation of the magnetic mean field
\begin{equation}
  \frac{\partial\left\langle \textbf{B}\right\rangle}{\partial t} = \mathbf{\nabla} \times \left\langle \textbf{v}\right\rangle \times \left\langle \textbf{B}\right\rangle -\eta_\mathrm{eff}\mathbf{\nabla} \times\mathbf{\nabla}\times\left\langle \textbf{B}\right\rangle
\label{meanB}
\end{equation}
with the effective parameter $\eta_\mathrm{eff}=\eta+T_\mathrm{L}(0)$. After a lengthy derivation \citep{BrandenburgSubramanian2005} this leads to
\begin{eqnarray}
  \frac{\partial M_\mathrm{L}}{\partial t} & = & 2\kappa_\mathrm{diff} M_\mathrm{L}'' + 2\left(\frac{4\kappa_\mathrm{diff}}{r}+ \kappa_\mathrm{diff}'\right) M_\mathrm{L}' + \frac{4}{r}\left(\frac{T_\mathrm{N}}{r} - \frac{T_\mathrm{L}}{r} - T_\mathrm{N}' - T_\mathrm{L}'\right) M_\mathrm{L} 
\label{dMLdt}
\end{eqnarray}
with
\begin{equation}
  \kappa_\mathrm{diff}(r) = \eta + T_\mathrm{L}(0) - T_\mathrm{L}(r).
\label{kappaN}
\end{equation}
The prime denotes differentiation with respect to $r$. The diffusion of the magnetic correlations, $\kappa_\mathrm{diff}$, contains in addition to the magnetic diffusivity $\eta$ the scale-dependent turbulent diffusion $T_\mathrm{L}(0) - T_\mathrm{L}(r)$.\\
In order to separate the time from the spatial coordinates we use the ansatz 
\begin{equation}
  M_\mathrm{L}(r,t) \equiv \frac{1}{r^2\sqrt{\kappa_\mathrm{diff}}}\psi(r)\mathrm{e}^{2\Gamma t}.
\end{equation}
Substitution of this ansatz in Eq.~(\ref{dMLdt}) gives us
\begin{equation}
  -\kappa_\mathrm{diff}(r)\frac{\mathrm{d}^2\psi(r)}{\mathrm{d}^2r} + U(r)\psi(r) = -\Gamma \psi(r).
\label{Kazantsev}
\end{equation}
This is the \textit{Kazantsev equation}, which is formally similar to the quantum-mechanical Schr\"odinger equation with a ``mass" $\hbar^2/(2\kappa_\mathrm{diff})$ and the ``potential"\footnote{We note that there is a typo in the paper of \cite{SchoberEtAl2012.1}, where the potential was derived for a general type of turbulence. The term $2\kappa_\mathrm{diff}$ / $r^2$ appeared here twice.}
\begin{equation}
  U(r) \equiv \frac{\kappa_\mathrm{diff}''}{2} - \frac{(\kappa_\mathrm{diff}')^2}{4\kappa_\mathrm{diff}} + \frac{2\kappa_\mathrm{diff}}{r^2} + \frac{2T_\mathrm{N}'}{r} + \frac{2(T_\mathrm{L}-T_\mathrm{N})}{r^2}. \\
\label{GeneralPotential}
\end{equation}
\begin{figure}[ht]
\begin{minipage}[b]{0.45\linewidth}
\centering
\includegraphics[width=\textwidth]{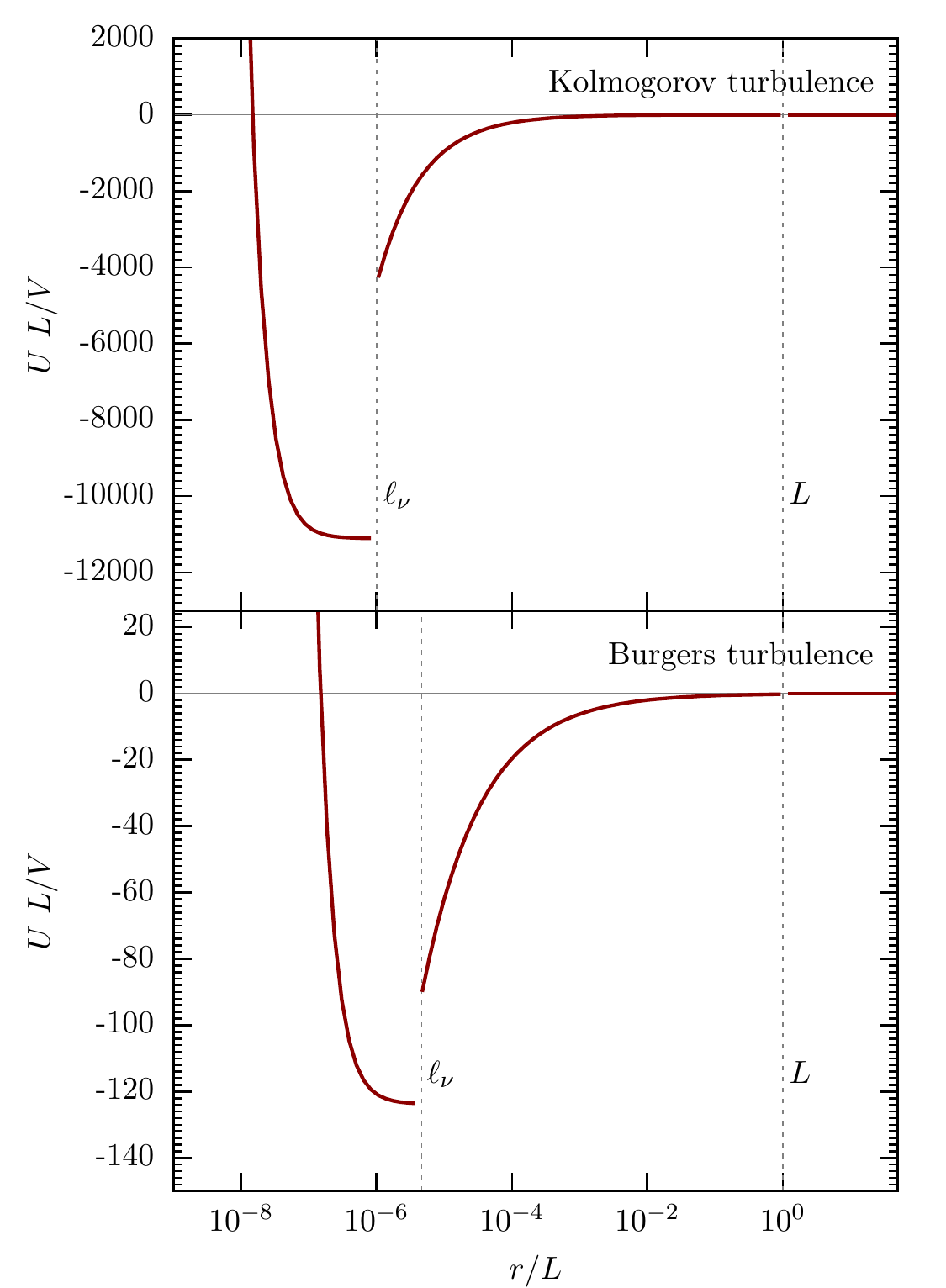} 
\caption{Normalized potential $U L/V$ as a function of normalized distance $r/L$. We choose here $\mathrm{Re}=10^8$ and $\mathrm{Rm}=10^{12}$ leading to $\mathrm{Pm}=10^4$.}
\label{Potential_Re1e8Rm1e12}
\end{minipage}
\hspace{0.5cm}
\begin{minipage}[b]{0.45\linewidth}
\centering
\includegraphics[width=\textwidth]{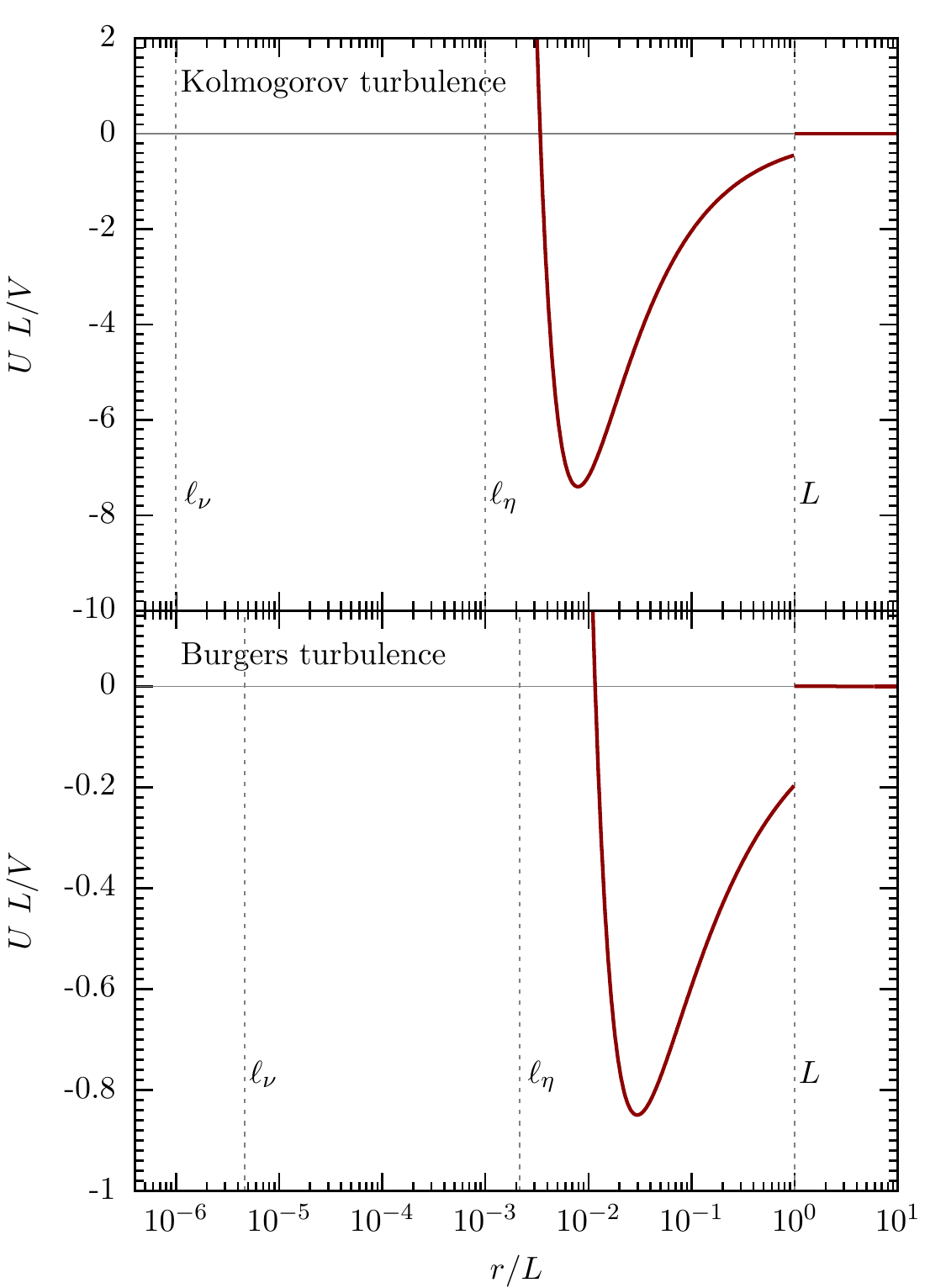} 
\caption{Normalized potential $U L/V$ as a function of normalized distance $r/L$. We choose here $\mathrm{Re}=10^8$ and $\mathrm{Rm}=10^{4}$ leading to $\mathrm{Pm}=10^{-4}$.}
\label{Potential_Re1e8Rm1e4}
\end{minipage}
\end{figure}
We solve the Kazantsev equation with the WKB approximation. For this we rewrite Eq.~(\ref{Kazantsev}) to
\begin{equation}
  \frac{\text{d}^2\theta(x)}{\text{d}x^2} + p(x)\theta(x) = 0
\label{Kazantsev2}
\end{equation}
with the substitutions $\psi(x) \equiv \text{e}^{x/2}\theta(x)$ and
\begin{equation}
  p(x) \equiv  - \frac{[\Gamma + U(x)]\text{e}^{2x}}{\kappa_\text{diff}(x)} - \frac{1}{4},
\label{p}
\end{equation}
where $x$ is a new coordinate with $r \equiv \text{e}^x$. Eq.~(\ref{Kazantsev2}) can be solved approximately by
\begin{equation}
  \int_{x_1}^{x_2}\sqrt{p(x')}\text{d}x' = \frac{\pi}{2},
\label{eigenvalue}
\end{equation}
where the integration limits are the zeros of $p(x)$. We have verified that the WKB approximation is valid in the limit of small and large magnetic Prandtl numbers.

\begin{figure}
 \includegraphics[width=3.4in]{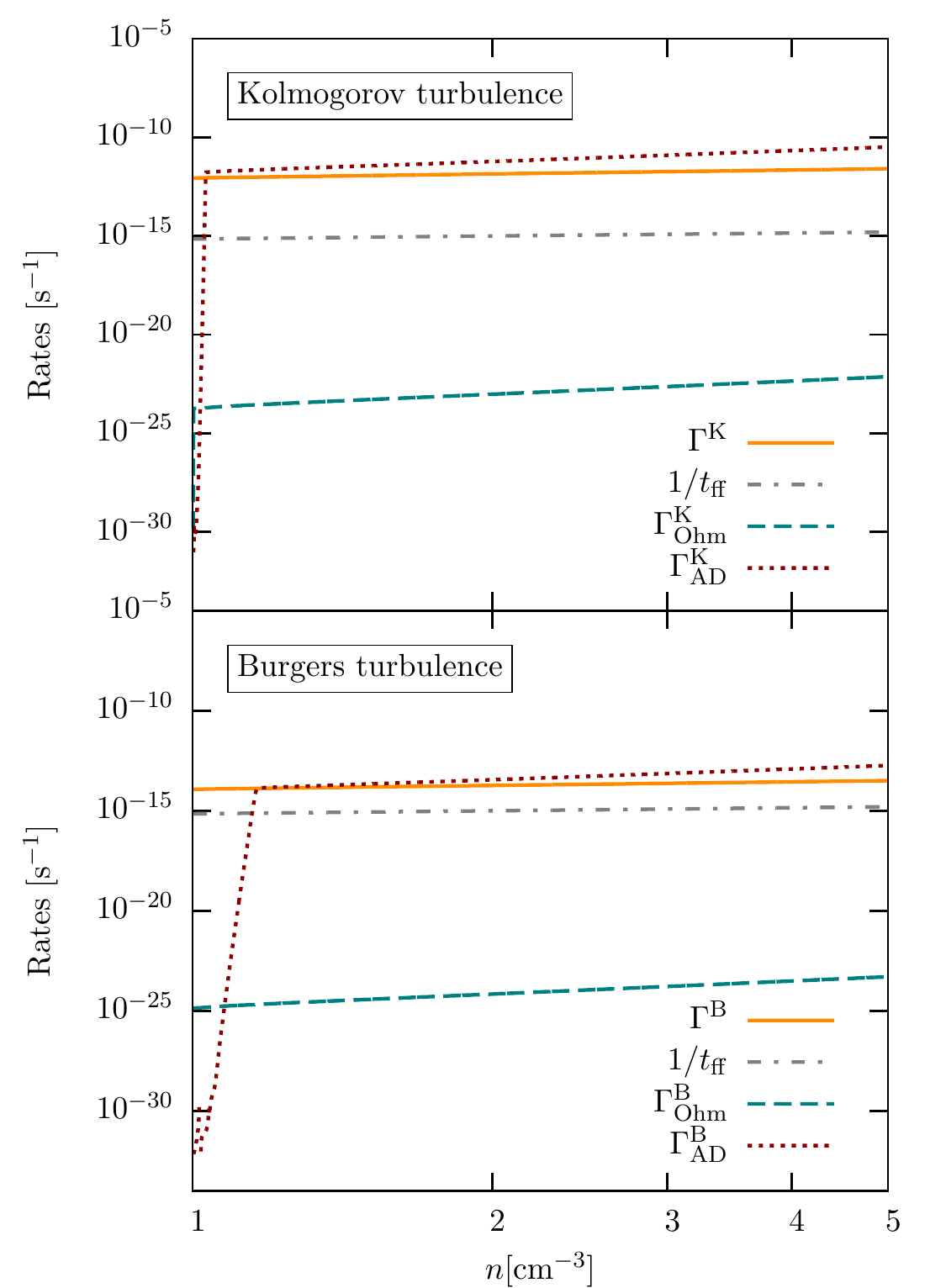} 
 \caption{The growth rate of the small-scale dynamo $\Gamma$ in the kinematic phase as a function of density $n$ compared to the inverse free-fall time $1/t_\mathrm{ff}$ and the diffusion rates due to Ohmic dissipation and ambipolar diffusion, $\Gamma_\mathrm{Ohm}$ and $\Gamma_\mathrm{AD}$.}
   \label{Rates}
\end{figure}

{\underline{\it Critical Magnetic Reynolds Numbers}}.
By setting the growth rate in our equations to zero and solving Eq.~(\ref{eigenvalue}) for Rm we calculate the critical magnetic Reynolds number $\text{Rm}_\text{crit}$ for small-scale dynamo action. We find that the small-scale dynamo is most efficient in the case of a purely rotational turbulent velocity field, i.e., for Kolmogorov turbulence, where $\text{Rm}_\text{crit}\approx110$. The critical magnetic Reynolds number for a turbulent field with vanishing rotational component, i.~e.~, Burgers turbulence, is roughly $2700$.\\
In the limit of large Pm the critical magnetic Reynolds number is not really a restriction, as the Re needs to be larger than about $10^3$ for turbulence. As here $\mathrm{Rm} \gg \mathrm{Re}$, Rm needs to be much larger than $10^3$ anyway. In the opposite limit of small Pm we have the condition of $\mathrm{Rm} \ll \mathrm{Re}$. For low Re, Rm can fall below $\mathrm{Rm}_\text{crit}$ and the small-scale dynamo eventually can not operate.

\begin{figure}[ht]
\begin{minipage}[b]{0.45\linewidth}
\centering
\includegraphics[width=\textwidth]{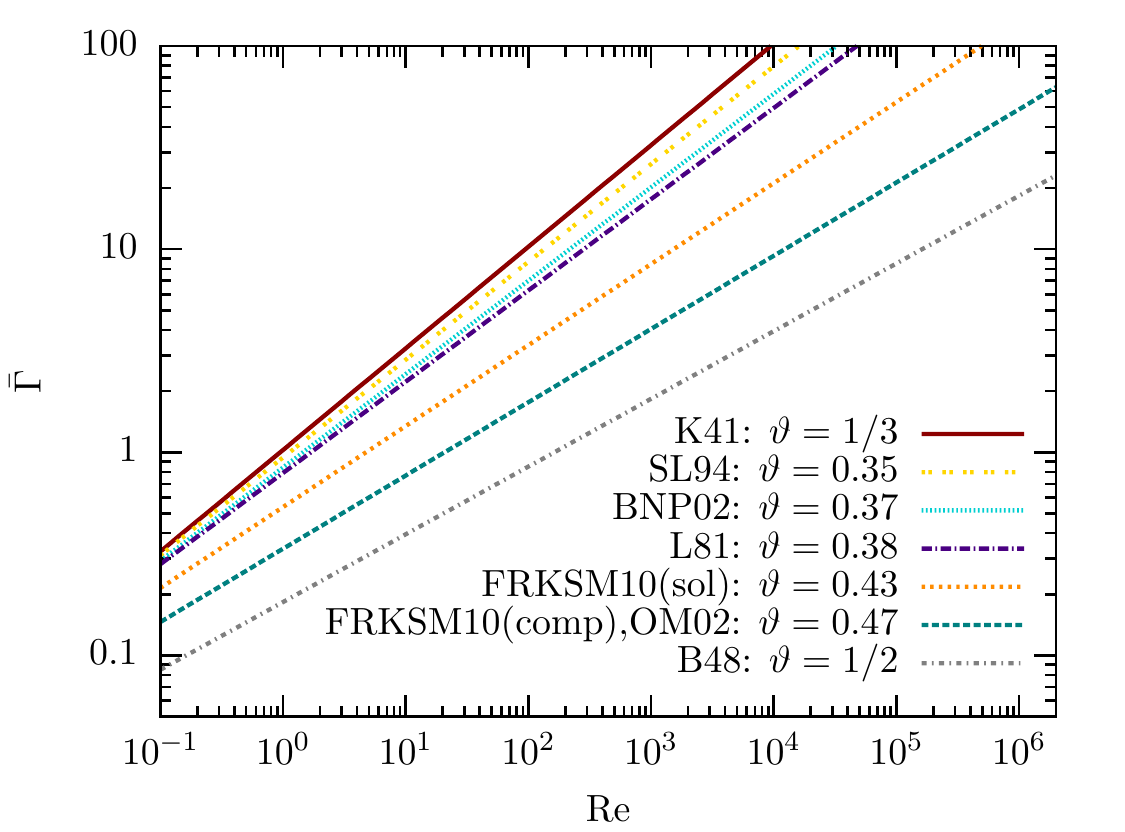}
\caption{The normalized growth rate $\bar{\Gamma}$ as a function of Re in the limit of large Pm. For $\vartheta$ we choose common values from literature: K41 \citep{Kolmogorov1941}, SL94 \citep{SheLeveque1994}, BNP02 \citep{BoldyrevEtAl2002}, L81 \citep{Larson1981}, FRKSM10 \citep{FederrathEtAl2010} (sol: solenoidal forcing; comp: compressive forcing), OM02 \citep{OssenkopfMacLow2002} and B48 \citep{Burgers1948}.}
\label{GrowthRate_largePm}
\end{minipage}
\hspace{0.5cm}
\begin{minipage}[b]{0.45\linewidth}
\centering
\includegraphics[width=\textwidth,height=0.825\textwidth]{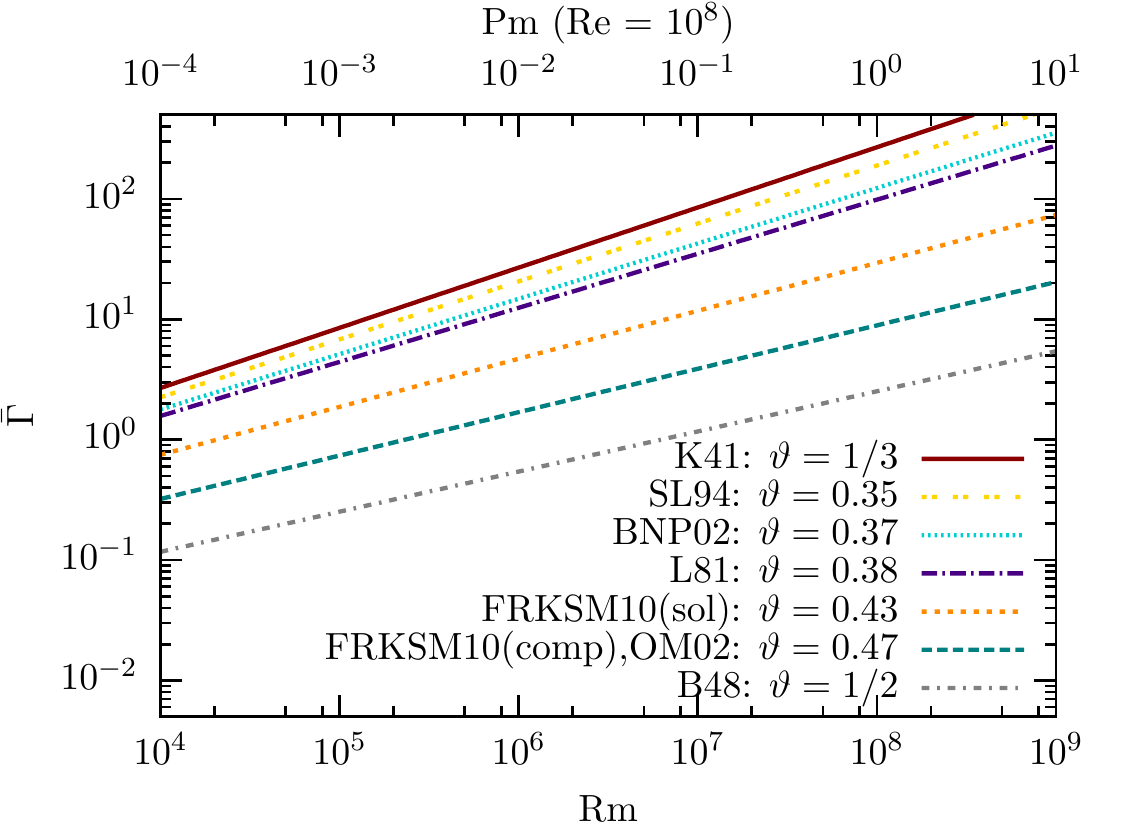}
\caption{The normalised growth rate $\bar{\Gamma}$ as a function of Rm (lower x axes) and Pm (upper x axes). The results shown for lower abscissa are only valid for small Pm, i.~e.~Rm $\ll$ Re, while we used a fixed Re of $10^8$ for the upper abscissa.}
\label{GrowthRate_lowPm}
\end{minipage}
\end{figure}

{\underline{\it Growth Rates}}.
In this paragraph we present general analytical solutions for the growth rate $\Gamma$ for an arbitrary slope of the turbulent velocity spectrum $\vartheta$, in the limits of large and low magnetic Prandtl numbers. \\
In the limit of large Pm the potential has its minimum at $\ell_\nu$ (see Fig.~\ref{Potential_Re1e8Rm1e12}). Thus, we expect the fastest growing mode to be in the viscous range, in which (\ref{p}) can be approximated as
\begin{equation}
  p(z) =  \frac{\text{Re}^{-(5+\vartheta)/(2+2\vartheta)}}{20} \frac{A_0z^2-B_0}{z^2}
\end{equation}
with the definitions
\begin{eqnarray}
  A_0 & = & \text{Re}^{(5+\vartheta)/(2+2\vartheta)}\left(163-304\vartheta\right)-\frac{20}{3} \text{Re}^{5/2} \bar{\Gamma}, \\
  B_0 & = & \left(304\vartheta-98\right)\text{Re}^2+\frac{20}{3}\text{Re}^{(2+8\vartheta)/(1+\vartheta)}\bar{\Gamma},
\end{eqnarray}
the new coordinate $z \equiv \left(\text{Re}^{3/2}\text{Pm}/3\right)^{1/2}y$ and the normalized growth rate $\bar{\Gamma} \equiv L/V~\Gamma$. The roots of $p(z)$ are $z_1=\sqrt{B_0/A_0}$ and $z_2=\sqrt{\text{Pm}/3}~\text{Re}^{(3\vartheta-3)/(4\vartheta+4)}$ leading to a solution of (\ref{eigenvalue}) in the limit of large Pm
\begin{eqnarray}
  \Gamma =  \frac{(163-304\vartheta)~V}{60~L}\text{Re}^{(1-\vartheta)/(1+\vartheta)}.
\label{Gamma}
\end{eqnarray}
In Fig.~\ref{GrowthRate_largePm} we show the dependency of $\bar{\Gamma}$ on the Re for different types of turbulence in the limit of large Pm. The growth rate ranges from $\Gamma\propto \text{Re}^{1/2}$ for Kolmogorov turbulence to $\text{Re}^{1/3}$ for Burgers turbulence. Altogether we find that the growth rate increases faster with increasing Re when the compressibility is lower.\\
The derivation of the growth rate is similar in the case of small Pm. However, the potential looks different in this limit (see Fig.~\ref{Potential_Re1e8Rm1e4}). The crucial discrepancy to the contrary limit of large Pm is that the potential only has a negative part between $\ell_\nu$ and $L$. Thus, there are only real positive eigenvalues of the Kazantsev equation (\ref{Kazantsev}) in the viscous range. The resulting equations in this case are more complicated, but with the ansatz
\begin{equation}
  \bar{\Gamma} = \alpha~\mathrm{Rm}^{\frac{1-\vartheta}{1+\vartheta}}
\label{ansatz}
\end{equation}
we can find an analytical solution of for the growth rate. This is motivated by our result in the limit of large Pm: $\bar{\Gamma} \propto Re^{(1-\vartheta)/(1+\vartheta)}$, where we replace Re by Rm, as the latter determines the characteristic scale for amplification $\ell_\eta$. Then the pre-factor of (\ref{ansatz}) turns out to be approximately
\begin{equation}
  \alpha =\frac{\vartheta\,(56-103\vartheta)}{5}\,a(\vartheta)^{\frac{\vartheta - 1}{1 + \vartheta }}\,\exp\left(\sqrt{\frac{5}{3\,\vartheta\,(56-103\vartheta)}}\,\pi \,\left(\vartheta - 1\right) - 2\right)
\end{equation}
with the abbreviation
\begin{eqnarray}
  a(\vartheta)  =  \frac{25 + {\sqrt{135\vartheta\,(56-103\vartheta) + {\left(\vartheta\,(79-157\vartheta) - 25\right)}^2}} - \vartheta\,(79-157\vartheta)}{\vartheta\,(56-103\vartheta)}.
\end{eqnarray}
We illustrate this result in Fig.\ \ref{GrowthRate_lowPm} for different types of turbulence.

{\underline{\it Evolution of the Magnetic Field on Small Scales}}.
In principle, the magnetic energy density, $E_\text{B} = B^2/(8\pi)$, evolves as 
\begin{equation}
  \frac{\text{d} E_\text{B}}{\text{d} t} = \left[\Gamma + \frac{4}{3 n} \frac{\text{d} n}{\text{d} t} - \Gamma_\text{Ohm} - \Gamma_\text{AD}(E_\text{B})\right] E_\text{B},
\label{BGrow}
\end{equation}
where we assume spherically symmetric collapse. By solving this equation numerically we find the evolution of the magnetic energy density on the viscous scale. As an initial field strength $B_0$ we use $10^{-20}~\text{G}$ on the viscous scale, which is a conservative value for a field generated by a Biermann battery \citep{Biermann1950,Xu2008}. In Fig.~\ref{MagneticFieldStrength} we show the resulting growth of the magnetic field strength. The field strength grows extremely rapidly as the density increases. We assume that the magnetic field is saturated as soon as equipartition with kinetic energy is reached.

\section{Non-Linear Small-Scale Dynamo}
\label{NonLinearSSD}

\begin{figure}
 \includegraphics[width=3.4in]{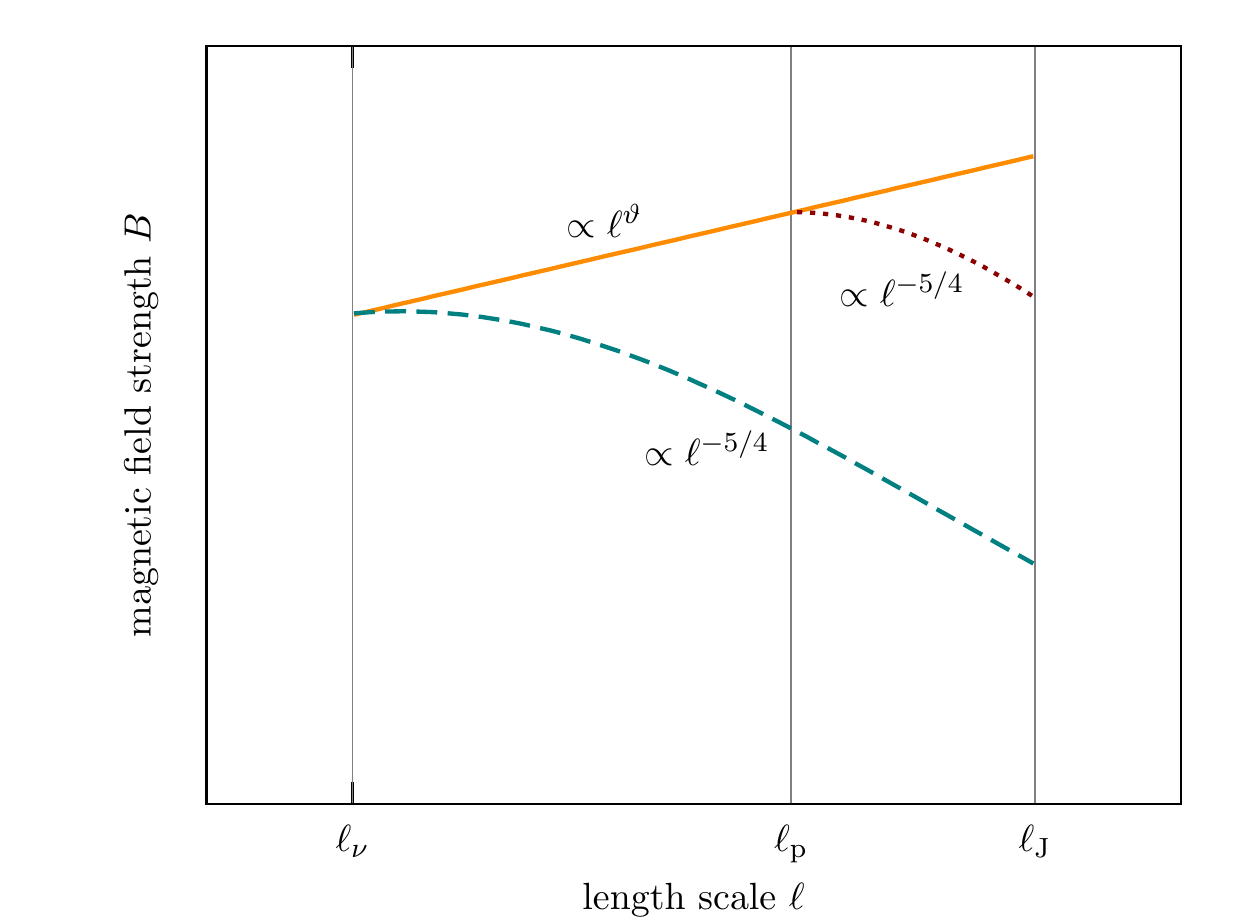} 
 \caption{Schematic model for the time evolution of the spectrum of the magnetic field strength in the inertial range of turbulence. For simplicity we use a  fixed frame of reference, where the viscous and the Jeans scale stay constant during the collapse. The different colors and line types represent the spectrum at different times. Further description can be found in the text.}
 \label{NonlinearDynamo}
\end{figure}
During saturation on the viscous scale, the coherence length of the magnetic field shifts towards larger scales, a well-known behavior for the small-scale dynamo \citep{SchekochihinEtAl2002,BrandenburgSubramanian2005}, recently shown to be true also in a collapsing system \citep{SurEtAl2012}. Analytical arguments suggest this occurs on the eddy-timescale of the current peak scale $\ell_\mathrm{p}$ leading to a time evolution
\begin{equation}
  \ell_\mathrm{p}(t) = \ell_\nu(t_\nu) + \left(\frac{v_\mathrm{J}}{\ell_\mathrm{J}^\vartheta} \left(t-t_{\nu}\right)\right)^{1/(1-\vartheta)},
\label{Shift}
\end{equation}
where $t_\nu$ is the point in time at which saturation occurs on the viscous scale.\\
We illustrate our model in Fig.~\ref{NonlinearDynamo}, where the different lines represent the spectrum at different times. The dashed green line is the spectrum at $t_\nu$, the dotted red line shows a later time and the solid orange line represents an even later point in time at which the magnetic field has saturated on the Jeans scale. The slope of the curves proportional to $\ell^{-5/4}$ is known as the Kazantsev slope in real space, which can be derived from the Fourier-transformed Kazantsev equation (\ref{Kazantsev}). The red line that connects the peak maxima at different times is a relic of the turbulence spectrum and thus is proportional to $\ell^{\vartheta}$. Saturation on a scale $\ell$ takes place, when the magnetic energy reaches equipartition with the turbulent kinetic energy. The saturated field strength is then
\begin{equation}
  B_{\ell,\mathrm{max}} = \sqrt{4 \pi \gamma k T n}~\left(\ell/\ell_\mathrm{J}\right)^{\vartheta}.
\label{Bmax}
\end{equation}
Using the Kazantsev slope, we can extrapolate the magnetic field strength onto the current Jeans length. By this we are able to determine the time evolution of the magnetic field on the Jeans scale.
\begin{figure}
  \includegraphics[width=3.4in]{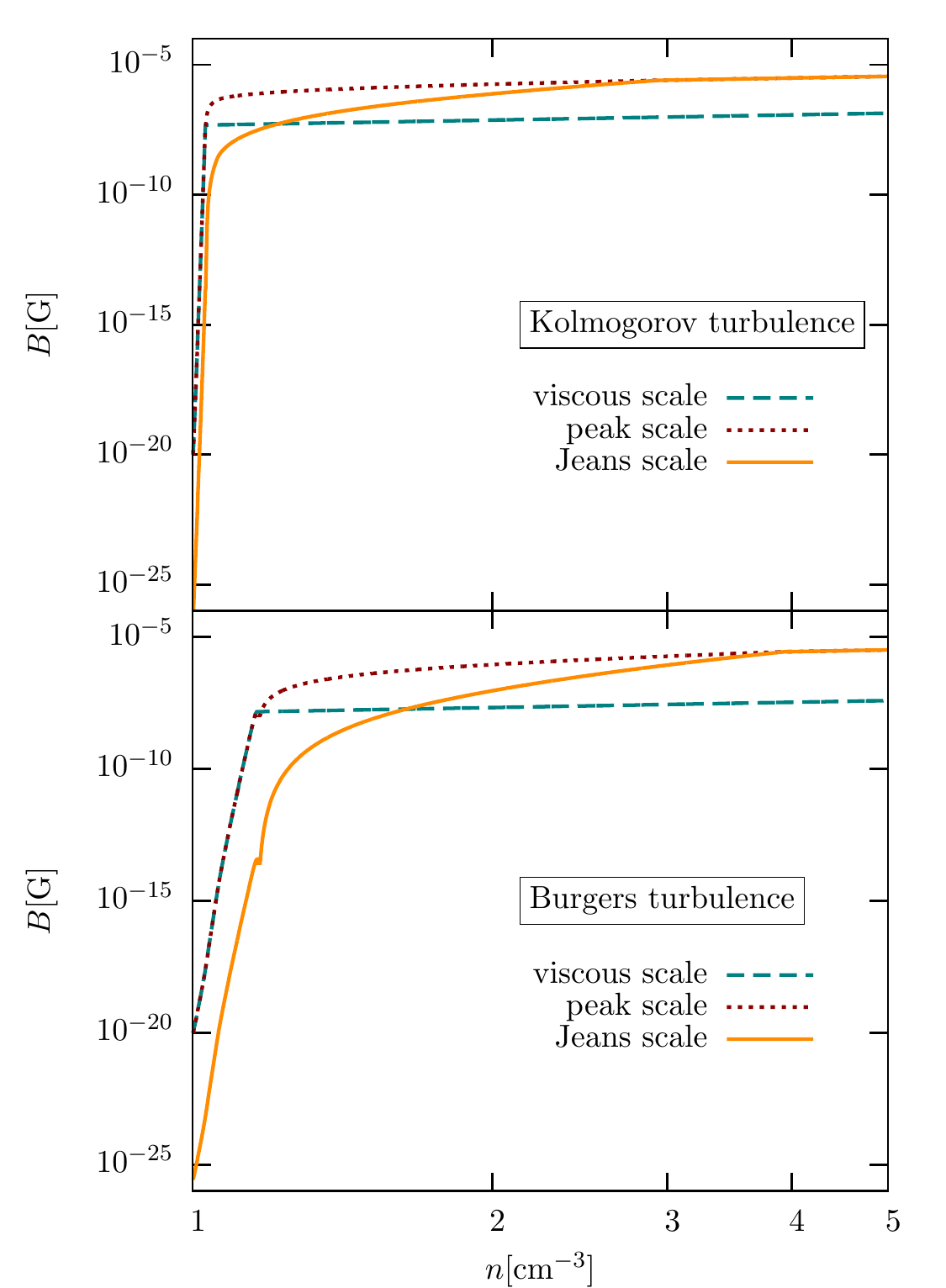} 
  \caption{The magnetic field strength as a function of the number density on the viscous scale (dashed green line), the peak scale (dotted red line) and the Jeans scale (solid orange line). The field approaches saturation extremely rapidly, when the density has only increased from $1 \mathrm{cm}^{-3}$ to $4 \mathrm{cm}^{-3}$ for all types of turbulence.}
  \label{MagneticFieldStrength}
\end{figure}
The result of the large-scale magnetic field is shown in Fig.~\ref{MagneticFieldStrength} together with the field on the current peak scale and the one on the viscous scale. One can see that the magnetic energy is shifted rapidly onto larger scales. For Kolmogorov turbulence the field on the Jeans scale saturates at a density of roughly 3 cm$^{-3}$ and for Burgers at a density of roughly 4 cm$^{-3}$. At the end of dynamo growth we have a magnetic field strength of about $10^{-6}$ G throughout the entire inertial range of the turbulence, i.~e.~within the Jeans volume.\\
After the rapid initial dynamo amplification the only way to amplify the magnetic field on the Jeans scale further is gravitational compression, which leads to $B \propto n^{2/3}$ for spherical collapse. However, the field has already reached equipartition with the kinetic energy at the end of dynamo amplification and, thus, increases only with $n^{1/2}$ (see Eq.~\ref{Bmax}).

\section{Conclusion}
\label{Conclusion}

We present a semi-analytical approach to the small-scale dynamo. Our starting point is the Kazantsev equation which we solve within the WKB approximation in the limit of large and small magnetic Prandtl numbers. We take into account the dependency on the type of turbulence, reaching from incompressible Kolmogorov turbulence to highly compressible Burgers turbulence.\\
We determine the critical magnetic Reynolds number, which needs to be exceeded for small-scale dynamo action. For Kolmogorov turbulence $\mathrm{Rm}_\mathrm{crit} \approx 110$, which is in good agreement with previous analytical and numerical results. For Burgers turbulence we find that $\mathrm{Rm}_\mathrm{crit} \approx 2700$. In the kinematic phase of the small-scale dynamo the magnetic energy grows exponentially with a growth rate $\Gamma$, which appears as an eigenvalue in the Kazantsev equation. We find in the limit of large Pm 
\begin{equation}
  \Gamma_{\mathrm{Pm}\gg1} \propto \text{Re}^{(1-\vartheta)/(1+\vartheta)}.
\end{equation}
Here the fastest growing mode is on the viscous scale, which is determined by Re. During the transition to low Pm, the resistive scale, determined by Rm, becomes larger than the viscous one and thus the amplification takes place at larger scales. For low Pm we find
\begin{equation}
  \Gamma_{\mathrm{Pm}\ll1} \propto \text{Rm}^{(1-\vartheta)/(1+\vartheta)}.
\end{equation}
We apply our theoretical model of the small-scale dynamo to the formation of the first stars. We model the chemistry and the thermodynamics during the collapse of a primordial halo and thus can determine the typical MHD quantities including Ohmic dissipation as well as ambipolar diffusion. We solve the energy equation and determine the evolution of the magnetic energy on the viscous scale. Furthermore, we model the nonlinear dynamo, where we assume that the magnetic energy is shifted to larger scales on the eddy timescale. By this we are able to follow the evolution of the magnetic field on the Jeans scale. During an increase of the density in the halo from $1 \mathrm{cm}^{-3}$ to $4 \mathrm{cm}^{-3}$ the dynamo saturates quickly in terms of density for all types of turbulence. We result in dynamically important magnetic fields of the order of a few $\mu\mathrm{G}$. \\
~\\
\small{\textbf{Acknowledgements}\\
{\small{We thank for funding through the {\em Deutsche Forschungsgemeinschaft} (DFG) in the {\em Schwer\-punkt\-programm} SPP 1573 ``Physics of the Interstellar Medium" (grant BA 3706/3-1, KL 1358/14-1, SCHL 1964/1-1) and SFB 881 ``The Milky Way System" (sub-projects B1 and B2). This project was also supported by the {\em Baden-W\"urttemberg-Stiftung} (P-LS-SPII/18). J.~S.~acknowledges support by IMPRS HD. D.~R.~G.~S.~thanks for funding via the SFB 963/1 on ``Astrophysical flow instabilities and turbulence". C.~F.~acknowledges funding provided by the {\em Australian Research Council} for an Australian Postdoctoral Fellowship under the Discovery Projects scheme (grant DP110102191). R.~B.~acknowledges funding from the DFG via SFB 676 ``Particles, Strings, and the Early Universe" (sub-project C9) and the Emmy-Noether grant BA 3706/1-1.}}

\end{document}